# SAC-CI Calculation of a Series of the Lowest $^2\Pi$-States of $HCl^+$ and $HBr^+$ Ions

Valerij S. Gurin[a], Mikhael V. Korolkov[b], Vitaly E. Matulis[c]

[a]Research Institute for Physical Chemical Problems, Belarusian State University, Minsk, Belarus
[b]Institute of Physics, NANB, Minsk, Belarus
[c]Belarusian State University, Minsk, Belarus

**Abstract**

The symmetry-adapted-cluster configuration interaction (SAC-CI) method is used for *ab initio* calculation of electronic structure of $HCl^+$ and $HBr^+$ molecular ions. Potential energy curves (PEC) are obtained for a series of low-lying $^2\Pi$-states starting from the ground state, $X^2\Pi$. The PEC for both $HCl^+$ and $HBr^+$ reveal the corresponding asymptotic behavior beginning from the interatomic distances ~8 $a_0$ for the first three states. These states are well separated, while a complex behavior occurs for the set of $4^2\Pi$-$6^2\Pi$ states with a bound character at short distances.

## 1. Introduction

Hydrogen halide molecules and molecular ions are very popular in molecular photochemistry as well in ultrafast laser spectroscopy [1-3] as examples of heterogeneous diatomics. High-level experiments on interaction of ultrashort light pulses with the molecules in controlled ground and excited states can be integrated with theory of the electronic structure of diatomics [4,5]. Calculations of electronic structure of the corresponding neutral molecules and molecular ions need taking into account many excited states, together with the data on the other molecular properties, dipole moments (DM), transition dipole moments (TDM) and vibrational spectra. In spite of great attention throughout years, there is a deficiency in knowledge of electronic states of neutral and ionic hydrogen halides, in particular $HCl^+$ and $HBr^+$. Meanwhile, a number of excited states (at least 4-5 states) need to describe adequately photochemistry of the molecules under powerful laser excitation. Within the framework of this study we consider two molecular ions from this series, $HCl^+$ and $HBr^+$. These molecular ions possess the doublet ground states, and there are no stable closed shell states. Open electronic shells dictate requirements to use high-level *ab initio* calculation methods with configuration interaction (CI). Up to date, there is no sufficient knowledge on a series of the states of $HCl^+$ and $HBr^+$ in the range of energies up to 5-6 eV above the ground state (i.e. the range of optical excitations). $HCl^+$ has been studied much more than $HBr^+$, however, the excited states above the 3rd one are worth to be recalculated and discussed more in detail. In our recent paper [6] we have presented a part of data for HCl$^+$ using the advanced *ab initio* calculation scheme, the method of symmetry-adapted-clusters configuration interaction (SAC-CI) [7,8]. The SAC-CI method has been developed for calculations of ground and excited states of molecular systems admitting a tuning of the calculation accuracy in correspondence with computing performance and system complicacy. The calculation scheme of SAC-CI takes an optimum cluster expansion for electronic excitations with selection of expansion terms based on symmetry of the system and has been shown to be successful for accurate description of different molecules [9,10]. In particular, the HCl ionization spectrum has been reproduced quite successfully [11] in the wide energy range. As compared with another multiconfiguration quantum chemical calculation approaches like CASSCF, CCSD, MRSDCI, etc., SAC-CI uses an optimized selection of the reference states to account CI of one- and multi-electron excitations and allows to attain good accuracy for many excited states.

As concerning all available calculation results for HCl$^+$ [6,11-14 and refs] together with electronic spectroscopy data [15-20], there is rather good consistence for the three states of $^2\Pi$-symmetry ($X^2\Pi$ is the ground state), three states of $^2\Sigma^+$-symmetry, while the other states are studied much less. In the case of HBr$^+$, the situation is else more complicated since calculations of molecules with Br atom requires a special accuracy in choice of basis sets, evaluation of spin-orbital interaction contributions, relativistic effects, etc. There are few reports [12,21-28] for the three first states of $^2\Pi$-symmetry, the two states of $^2\Sigma^+$-symmetry, the lowest states of $^2\Sigma^-$- and $^2\Delta$-symmetry. Two lowest quartet states $^4\Pi$ and $^4\Sigma^-$ were also calculated, and they enter this energy range. Thus, a more detailed study for a series of electronic states in the lowest energy range (~12 eV) need for understanding the electronic structure of HBr$^+$. Within the framework of this publication we have restricted ourselves to a series of $^2\Pi$-states from the calculation data without accounting the spin-orbital effects as the first approximation comparing with the similar calculations of HCl$^+$. The ground state for both ions is $X^2\Pi$, and the $^2\Pi$-series is to be responsible for photoexcitation processes from this state. The states of other symmetries are the subject of subsequent works.

## 2. Calculation details

The SAC-CI method was used for electronic structure calculations of both HCl$^+$ and HBr$^+$ at the general-R level including R-operators up to the third order, which takes into account the excitations including sextuplet states.

We use the all-electron basis sets with additional diffuse components justified by previous calculations and our tests have provided good accuracy for experimental spectroscopic constants corresponding to ground states of $HCl^+$ and $HBr^+$ ($X^2\Pi$ state for both). For Cl atom two diffuse s- and p-functions with exponents $\xi_s$=0.06, 0.015 and $\xi_p$=0.04, 0.01 and four d-functions with $\xi_d$=0.7, 0.25, 0.08 and 0.02 were added to the standard 6-311G basis [29] adopted for calculation of the experimental ionization spectrum of HCl. The values of exponents were shown to be the optimum for Cl and H atoms [30]. For Br atom we have chosen the basis set aug-cc-pVDZ completing a series of test calculations. This is a correlation-consistent basis set providing good results in high-level calculations with CI, including heavy main group elements. For hydrogen atoms the basis set of 6-311G quality was used with two p-functions with $\xi_p$=1.0 and 0.3.

In the CI procedure for calculation of the both molecular ions, the external electronic shell was treated as active orbitals: in the case of $HCl^+$ 1s, 2s and 2p atomic orbitals of Cl were used as frozen and in the case of $HBr^+$ 1s, 2s, 2p, 3s, 3p orbitals of Br were frozen. Thus, the active space for CI included the valence orbitals beginning from and 1s- for H, 3s- for Cl, 4s- for Br consisting of 4 occupied and 58 virtual orbitals. This choice of active orbitals is typical also for different molecules with n$p$-elements. All calculations were carried out with Gaussian 03 software [31].

## 3. Calculation results

A series of PEC for the six lowest doublet $^2\Pi$-states is presented in Fig. 1. They are rather similar for $HCl^+$ and $HBr^+$. For the first three states recently [6] (including the comparing them with previous data [14]) we have demonstrate quite good consistence of the results derived within the various calculation methods evidencing that principal contributions of the electronic correlation have been taken into account correctly in the different approaches under study. The next effects, untreated yet (spin-orbital, relativistic, etc.) are of minor importance for $HCl^+$, but need to be treated for $HBr^+$ to obtain a splitted set of $^2\Pi$-states. The present comparison is given with no spin-orbital splitting. Table 1 collects the numerical characteristics for the dissociation paths of these six $^2\Pi$-states compared with corresponding reference data.

Among these six states the lowest one, ground state, $X\,^2\Pi$, is bound for both $HCl^+$ and $HBr^+$ and the other two states in this energy range are unbound. The bound character is observed also for $4^2\Pi$, $5^2\Pi$ and $6^2\Pi$ at short distances, and this fact is very similar for the both. There are different dissociation channels: for the ground states, $X\,^2\Pi$, the channel with ionized halogens, $H^0 + X^+$, appears for the both ions, while, the next states, $2^2\Pi$ and $3^2\Pi$ demonstrate different dissociation channels for $HCl^+$ and $HBr^+$. $4^2\Pi$-$6^2\Pi$ series result in $H^+ + X^0$ for both cases. Thus, the behavior of $2^2\Pi$ and $3^2\Pi$ states present the principal difference for the ions at the level theory used. A difference for $4^2\Pi$-$6^2\Pi$ can appear under more detailed consideration of avoiding crossing effects since they are very close in energies with non-monotonous energy variations with distance. The values of energy of corresponding distant atoms and ions for all states in Tables 1,2 indicate that the sequence of atomic terms is correctly reproduced, though the absolute values deviate in some cases from the experimental reference values.

$4^2\Pi$–$6^2\Pi$ states enter the noticeably higher energy interval, ~ 10 eV higher than the ground one. This separation is common feature both for HCl$^+$ and HBr$^+$. Under analysis of photodissociation dynamics they should be taken into account if the high-energy excitation is used, but at the first approximation in the visible and near-UV excitations, the states beginning from $4^2\Pi$ may be neglected. The significant energy separation of $4^2\Pi$–$6^2\Pi$ and $X^2\Pi$–$3^2\Pi$ states can be conditioned by the strongly elevated energies of the corresponding atomic orbitals contributing to these sets of states. The first three states appear mainly due to contributions of *3p-* (for HCl$^+$) and *4p-* (for HBr$^+$) orbitals while the electronic configuration of the higher states include essential *3s-*, and *4s-* contributions, respectively. Such structure of orbitals in molecular ions can provide close data in energies of $4^2\Pi$–$6^2\Pi$ states and the more variation of energies of the three lowest states.

The calculations done for the higher $^2\Pi$-states (Fig. 1,2) indicate also at least one minimum at the short distances, near $R_e$, for each state, and at the larger R, a non-monotonous behavior is seen that can be a result of the avoiding crossing effects for close states. The points of possible avoiding crossing appear almost at the same distances for HCl$^+$ and HBr$^+$. It should be noticed that the higher states do not demonstrate good asymptotic behavior even up to distances R ~ 20 $a_0$ while the lowest three states starting from R ~ 8-9 $a_0$ have unchanged energies.

## 4. Conclusions

Quantum chemical calculations with the SAC-CI method have been used for study of a series of doublet $^2\Pi$ electronic states of HCl$^+$ and HBr$^+$. The calculation data for PEC of HCl$^+$ and HBr$^+$ are rather similar within the framework of the given calculation level (without spin-orbital contributions). The calculated PECs reproduce well the data for HCl$^+$ and HBr$^+$ doublet states calculated earlier and demonstrate new features for a series of the higher states. At the large interatomic distances the PECs correlate with different dissociation channels of the correct asymptotic ionic compositions.

Table 1. Calculated data for dissociation paths and asymptotic separations for six lowest $^2\Pi$ states of HCl$^+$

| State of HCl$^+$ | Dissociation limits and their atomic terms | Energy at the distance R=20a$_0$, Hartree | Calculated energies, eV | Reference experimental energies, eV [32] |
|---|---|---|---|---|
| $X^2\Pi$ | H$^0$($^2$S)+Cl$^+$($^3$P) | -459.64983 | 0 | 0 |
| $2^2\Pi$ | H$^+$+Cl$^0$($^2$P) | -459.61832 | 0.857 | 0.6308 |
| $3^2\Pi$ | H$^0$($^2$S)+Cl$^+$($^1$D) | -459.59135 | 1.591 | 1.445 |
| $4^2\Pi$ | H$^+$+Cl$^0$($^2$P) | -459.28695 | 9.88 | 9.87 |
| $5^2\Pi$ | H$^+$+Cl$^0$($^2$D) | -459.24401 | 11.04 | 11.06 |
| $6^2\Pi$ | H$^+$+Cl$^0$($^2$D) | -459.236849 | 11.24 | 11.14 |

Table 2. Calculated data for dissociation paths and asymptotic separations for six lowest $^2\Pi$ states of HBr$^+$

| State of HBr$^+$ | Dissociation limits and their atomic terms | Energy at the distance R=20a$_0$, Hartree | Calculated energies, eV | Reference experimental energies, eV [32] |
|---|---|---|---|---|
| $X^2\Pi$ | H$^0$($^2$S)+Br$^+$($^3$P) | -2572.565627 | 0 | 0 |
| $2^2\Pi$ | H$^0$($^2$S)+Br$^+$($^1$D) | -2572.514556 | 1.399 | 1.499 |
| $3^2\Pi$ | H$^+$+Br$^0$($^2$P) | -2572.493161 | 1.972 | 1.785 |
| $4^2\Pi$ | H$^+$+Br$^0$($^2$P) | -2572.188107 | 10.27 | 10.22 |
| $5^2\Pi$ | H$^+$+Br$^0$($^2$D) | -2572.165753 | 10.88 | 11.19 |
| $6^2\Pi$ | H$^+$+Br$^0$($^2$D) | -2572.137068 | 11.66 | 11.67 |

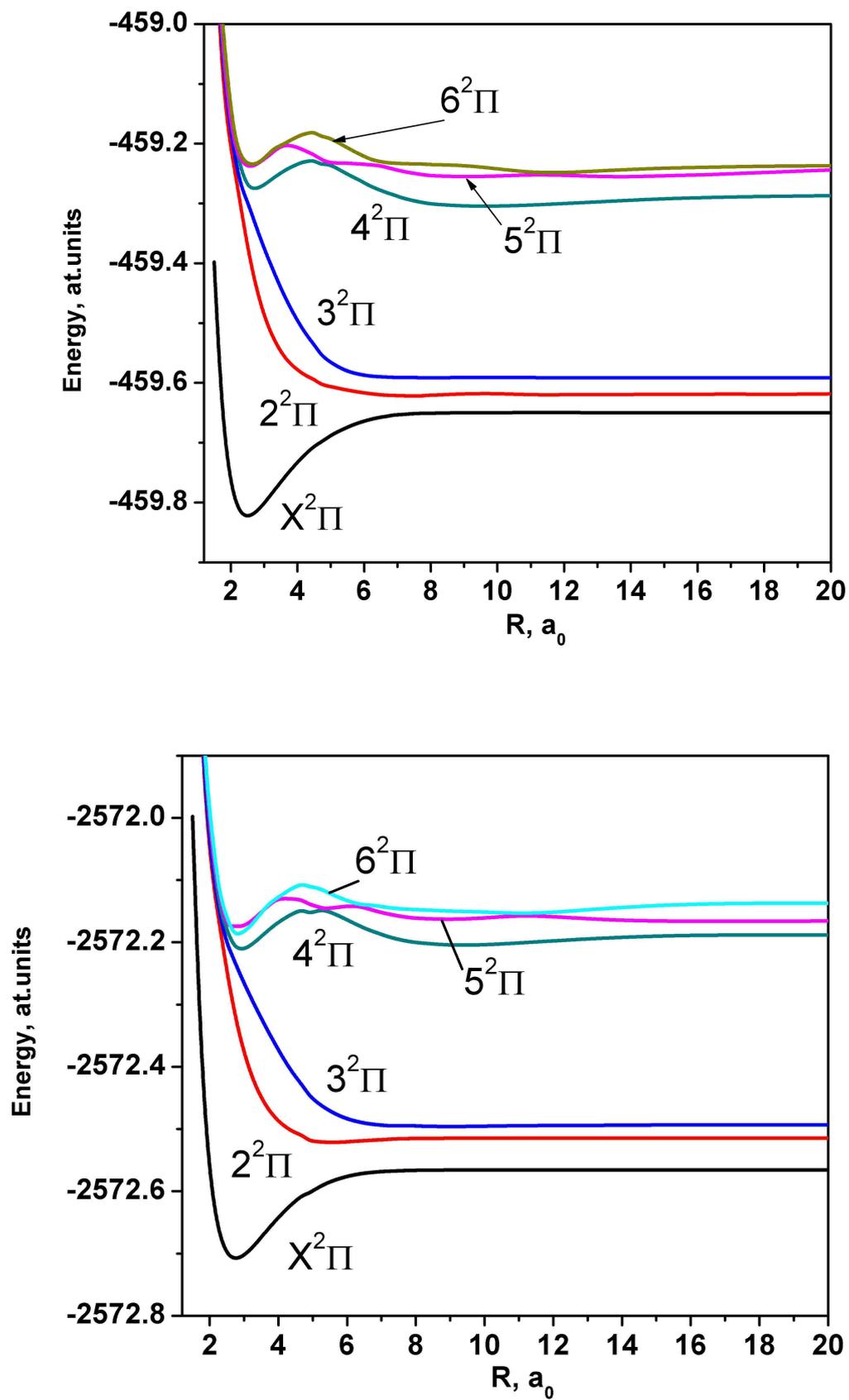

Fig. 1. Potential energy curves for a series of $^2\Pi$ states of HCl$^+$ and HBr$^+$ (upper and lower parts, respectively).